\documentclass[pdftex,twocolumn,epjc3]{svjour3}      
\RequirePackage{amsmath}

\newcommand{\be}{\begin{equation}}
\newcommand{\ee}{\end{equation}}
\newcommand{\ba}{\begin{eqnarray}}
\newcommand{\ea}{\end{eqnarray}}
\newcommand{\nn}{\nonumber\\}

\RequirePackage{graphicx}
\RequirePackage{mathptmx}      
\RequirePackage{flushend}
\RequirePackage[numbers,sort&compress]{natbib}
\RequirePackage[colorlinks,citecolor=blue,urlcolor=blue,linkcolor=blue]{hyperref}

\begin{document}
		\title{Energy-loss of heavy quarks in the isotropic collisional hot QCD medium at a finite  chemical potential}
	
	\author{M. Yousuf Jamal\thanksref{e1,addr1}	\and
		Bedangadas Mohanty\thanksref{addr1} }
	
	\thankstext{e1}{mohammad.yousuf@niser.ac.in}
			\institute{School of Physical Sciences, National Institute of Science Education and Research, HBNI, Jatni-752050, India \label{addr1}}
		\maketitle

\begin{abstract}
The present article is the followup of {\color{blue} Eur.\ Phys.\ J.\ C {\bf 79}, 761 (2019)}, where we have studied the energy-loss of the heavy quarks traversing through the isotropic collisional hot QCD medium. Since, the exploration of QCD phase diagram is possible with the upcoming experimental facilities such as Anti-proton and Ion Research (FAIR) and Nuclotron-based Ion Collider fAcility (NICA), at finite baryon density and moderate temperature, the inclusion of finite chemical potential is essential to study the hot QCD/QGP medium. Therefore, the aim is to develop a formalism to study the energy-loss of heavy quarks moving in the interacting collisional hot QCD medium having small but a finite quark chemical potential. To do so, the extended effective fugacity quasi-particle model has been employed ~\cite{chandra_quasi1, chandra_quasi2, Mitra:2017sjo} while considering the effective kinetic theory approach using the Bhatnagar-Gross-Krook (BGK) collisional kernel. Finally, the momentum dependence of the energy-loss for the charm and bottom quark has been investigated at different values of collisions frequency and chemical potential. It is observed that  as compared to charm quark, bottom quark loses less energy at a particular momentum, collisional frequency and chemical potential. Also, the energy-loss is seen to decrease with increasing chemical potential. 
\end{abstract}

	\section{Introduction}
	The exploration of the QCD phase diagram is an important part of modern high-energy physics (HEP). The different phases of the QCD phase diagram play an important role in the study of the early universe and also the core of neutron stars. Relativistic heavy-ion collision (HIC) experiments provide a unique opportunity of creating such phases in the laboratory. The upcoming experimental facilities such as FAIR, NICA, {\it etc}, are supposed to explore the QCD phase diagrams at a finite chemical potential. Therefore, the phase transition from hadronic to partonic degrees of freedom at finite baryon densities is one among the most interesting challenges of relativistic HIC  and hence, the theoretical estimates/studies need the consideration of finite chemical potential. The main problem in the study of the medium produced at various experimental facilities is its small size and short-lived nature. Therefore, to study the created hot QCD medium, one needs to trace back using information observed at the detector end (as particle spectra).
	The heavy quarks, on the other hand, due to their slow thermalization act as independent degrees of freedom and provide an opportunity to study the medium created in HIC experiments. While passing through the interacting hot QCD medium created after the collision, the heavy quark loses its energy and at the end causes suppression in the yield of high $p_T$ hadrons.

The energy-loss of heavy quarks moving in the hot QCD medium have been studied by many researchers and several excellent articles are present in the literature. Earlier, the collisional energy-loss suffered by the high energy partons due to the elastic scatterings off thermal quarks and gluons in QCD plasma was studied by Bjorken~\cite{bjorken1982energy}. Later, a formalism was developed by Thoma and Gyulassy~\cite{Thoma:1990fm} in which they obtained the collisional energy-loss in terms of the longitudinal and transverse dielectric functions. Braaten and Thoma ~\cite{Braaten:1991jj} had constructed a systematic framework of the energy-loss for both soft and hard momentum transfers within the finite temperature field theory approach ~\cite{Mrowczynski:1991da, Thomas:1991ea, Koike:1992xs}. The anisotropic effects in the context of heavy quarks energy-loss have been studied in Refs.~\cite{Romatschke:2004au, Baier:2008js, Carrington:2015xca}. Apart from these, there are several informative articles in which the authors have discussed the energy-loss of heavy quarks either through radiative or collisional means~\cite{Baier:2000mf, Jacobs:2004qv, Armesto:2011ht, Majumder:2010qh, Mustafa:1997pm, Dokshitzer:2001zm, Djordjevic:2003zk, Wicks:2007am, Abir:2011jb, Jeon:2003gi, Djordjevic:2008iz,    
	Gyulassy:1999zd, Zakharov:2000iz,  Baier:2001yt, Qin:2007rn, Cao:2013ita, Mustafa:2003vh, DuttMazumder:2004xk, Meistrenko:2012ju,
	Burke:2013yra, Peigne:2007sd, Neufeld:2014yaa, Chakraborty:2006db, Adil:2006ei, Peigne:2005rk, Dusling:2009jn, Cho:2009ze}.
The polarization energy-loss of heavy quarks while considering the hot viscous QGP has been studied recently by Jiang et al.~\cite{Jiang:2014oxa, Jiang:2016duz}.
Considering the transport theory approach the same while employing the relaxation time (RTA) collisional kernel has been studied in Ref.~\cite{Elias:2014hua} while on the other hand with the Bhatnagar-Gross-Krook (BGK) collisional kernel in Ref.~\cite{Han:2017nfz}. Considering the ADS/CFT approach, the energy-loss of moving heavy quarks has been studied in Refs.~\cite{Fadafan:2008gb, Fadafan:2008uv, Fadafan:2012qu}.

In our previous work done in Ref. ~\cite{Jamal:2019svc}, we studied the energy-loss of heavy quark passing through the interacting collisional hot QCD medium. There we adopted the effective kinetic theory approach and employed the BGK as well as RTA kernels at vanishing chemical potential. To incorporate the hot QCD medium interaction effects, we employed the effective fugacity quasi-particle model ~\cite{chandra_quasi1, chandra_quasi2}. 
In the present article, the aim is to study the energy-loss of heavy quarks (charm and bottom) moving in the interacting collisional hot QCD/QGP medium having small but finite quark chemical potential. To do so, again the effective kinetic theory approach while considering BGK- collisional kernel has been employed along with the extended EQPM.

The manuscript is organized as follows. In section~\ref{sec:EL}, we shall provide the formalism for the energy-loss of heavy quarks moving in the interacting collisional hot QCD medium having finite chemical potential. In sections~\ref{el:RaD}, we shall discuss the various results of the energy-loss at different values of chemical potential and collision frequency. Section~\ref{el:SaF}, will have the summary and future possibilities of the present work.	
	
\section{Formalism for the energy-loss of moving heavy quarks at finite chemical potential}
\label{sec:EL}
The heavy quark while passing through the hot QCD medium induces the chromoelectric field and generates the color Lorentz force that acts against its motion. It causes the energy-loss of the heavy quark and slows down its speed. In this case, the energy-loss per unit length is given as \cite{Jamal:2019svc, Thoma:1990fm, Koike:1992xs, Elias:2014hua, Han:2017nfz},  
\ba
-\frac{\mathrm{dE}}{\text{d{\bf x}}}=g_s~q^a~\frac{{\bf u}}{|{\bf u}|}\cdot \mathrm{{\bf E}_{ind}^{a}}(X).
\label{eq:el}
\ea
where $g_s$ is the strong coupling constant and $q^a$ are the quarks color charges. The velocity is given as, ${\bf u}=\frac{\bf p}{\sqrt{{\bf p}^2+m^2}}$ ( $m$ being the mass of the heavy quark and ${\bf p}$ the momentum vector). Here, we have $\mathrm{N_{c}^{2}-1}$ induced chromo-electric fields which belong to the $\mathrm{SU(N_c)}$ gauge group. Eq.(\ref{eq:el}) can be further written in terms of the longitudinal and transverse components of dielectric permittivity of the isotropic hot QCD/QGP  medium having finite chemical potential and temperature as \cite{Jamal:2019svc},
\ba
	-\frac{\text{dE}(\mu_{q}, T)}{\text{d {\bf x}}}&=&\frac{C_F \alpha _s (\mu_{q}, T) }{2 \pi ^2 |{\bf u}|}\int^{k_\infty}_{k_0}\omega d\Omega dk\bigg\{\left(k^2 |{\bf u}|^2-\omega ^2\right)\nn &\times&\text{Im}(\omega ^2 \epsilon_T(\mu_{q}, T)-k^2)^{-1}\nn 
	&+&\text{Im}\epsilon^{-1}_{L}(\mu_{q}, T)\bigg\}_{\omega ={\bf k \cdot u}},
	\label{eq:de}
\ea
 where, $C_F =4/3$ is the Casimir invariant in the fundamental representation of the $\mathrm{SU(3)}$. The QCD running coupling constant, $\alpha_{s}(\mu_{q}, T)$ at finite chemical potential and temperature~\cite{ Mitra:2017sjo, Srivastava:2010xa} given as,
\ba
\alpha_{s}(\mu_{q}, T)&=&\frac{g^2_{s}(\mu_{q}, T)}{4 \pi}\nn
&=& \frac{6\pi}{\left(33-2N_f\right)\ln\left(\frac{T}{\Lambda_T} \sqrt{1+\frac{\mu^2_q}{\pi^2 T^2}}\right)}\nn
&\times& \bigg(1-\frac{3\left(153-19N_f\right)}{\left(33-2N_f\right)^2}
\frac{\ln \left(2\ln \left( \frac{T}{\Lambda_T} \sqrt{1+\frac{\mu^2_q}{\pi^2 T^2}}\right)\right)}{\ln \left( \frac{T}{\Lambda_T} \sqrt{1+\frac{\mu^2_q}{\pi^2 T^2}}\right)}\bigg).\nn
\ea
Next, to obtain the longitudinal and transverse part of the dielectric permittivity,  we first obtain the current induced inside the medium, $J_{ind,a}^{\mu}(X)$  due to the change in the particle distribution function given as~ \cite{Mrowczynski:1993qm, Romatschke:2003ms, Jiang:2016dkf,Schenke:2006xu},
\ba
J_{ind,a}^{\mu}(X)&=&g\int\frac{d^{3}p}{(2\pi)^3} u^{\mu}\{2N_c \delta f^{g}_a(p,X)+N_{f}[\delta f^{q}_a(p,X) \nn
&-&\delta f^{\bar{q}}_a(p,X)]\}\label{indcurrent}.
\ea
Where the change in the distribution functions, $\delta f^{i}$ can be obtained by solving the Boltzman-Vlasov transport equation that can be  written for each individual species ($i\in$ \{quarks, anti-quarks and gluons\}) as,
\ba
u^{\mu}\partial_{\mu} \delta f^{i}_a(p,X) + g \theta_{i}
u_{\mu}F^{\mu\nu}_a(X)\partial_{\nu}^{(p)}f^{i}(\mathbf{p})=C^{i}_a(p,X),    \label{transportequation}
\ea
where, $\theta_{i}\in\{\theta_g,\theta_q,\theta_{\bar{q}}\}$ and have the values $\theta_{g}=\theta_{q}=1$ and $\theta_{\bar{q}}=-1$.   
The partial four derivatives, $\partial_{\mu}$, $\partial_{\nu}^{(p)}$ correspond to the space and momentum,
respectively. The collisional  kernel, ${C}^{i}_a(p,X)$ is considered here to be the BGK-type  ~\cite{Bhatnagar:1954} given as follows, 
\ba
{C}^{i}_a(p,X)=-\nu\left[f^{i}_a(p,X)-\frac{N^{i}_a(X)}{N^{i}_{\text{eq}}}f^{i}_{\text{eq}}(|\mathbf{p}|)\right]\,\text{,}\label{collision}
\ea
where, 
\ba f^{i}_a(p,X)=f^{i}(\mathbf{p})+\delta f^{i}_a(p,X),
\ea
are the distribution functions of
quarks, anti-quarks and gluons, $f^{i}(\mathbf{p})$ is
equilibrium part while, $\delta f^{i}_a(p,X)$
is the perturbed part of the distribution functions such that $\delta f^{i}_a(p,X)\ll f^{i}(\mathbf{p})$.
$N^{i}_a(X)$ is the particle number and $N^{i}_{\text{eq}}$ is it value at equilibrium.  The collisions frequency,  $\nu$ is  considered here to be independent of momentum and particle species. The self energy, $\Pi^{\mu\nu}(K)$ in the Fourier space, can be obtained from the induce current as,
	\ba
	\Pi^{\mu\nu}(K)=\frac{\delta J^{\mu}_{a, ind}(K)}{\delta A_{\nu, a}(K)},
\label{eq:p}	 
	\ea
	where, $K\equiv K^{\mu} =(\omega,{\bf k })$. The dielectric permittivity, $\epsilon^{ij}$ in the temporal axial gauge can be obtained from the gluon selfenergy, $\Pi^{ij}$ as, 
	\ba
		\epsilon^{ij}=\delta^{ij}-\frac{1}{\omega^2}\Pi ^{\text{ij}}.
		\label{eq:ep}
\ea
Solving Eq.~(\ref{indcurrent}) in the Fourier space along with Eq.~(\ref{transportequation}) and Eq.~(\ref{eq:p}), we obtained the spatial component of self-energy as,
\ba
\Pi ^{ij}(K,\nu, \mu_q,T)&=& m_D^2(T,\mu_q)\int \frac{d\Omega }{4 \pi }u^{i} u^{l}\Big\{u^{j} k^{l}\nn
&+&\left(\omega -{\bf k}\cdot {\bf u}\right)\delta ^{lj}\Big\} D^{-1}\left(K,\nu \right),
\label{eq:pi}
\ea
where,  $D\left(K,\nu \right)=\omega +i \nu-{\bf k}\cdot {\bf u}$. The  squared Debye mass  given as,
\ba
m^{2}_D(T,\mu_q) &=& 4\pi \alpha_s(T,\mu_q) \bigg(-2N_c \int \frac{d^3 p}{(2\pi)^3} \partial_p f_g(p)\nn &-& N_f \int \frac{d^3 p}{(2\pi)^3} \partial_p \left(f_q(p)+f_{\bar q}(p)\right)\bigg).
\label{eq:md} 
\ea    
To solve Eq.~(\ref{eq:pi}) analytically, the $\Pi^{ij}$ is further decomposed in terms of its longitudinal and transverse parts as,
\ba
\Pi^{ij}(K,\nu,T,\mu_q)= A^{ij}~\Pi_T(K,\nu,T,\mu_q)+ B^{ij}\Pi_L(K,\nu,T,\mu_q),
\label{eq:eplt}\nn
\ea
where, $A^{\text{ij}}=\delta ^{\text{ij}}-\frac{k^i k^j}{k^2}$, and $ B^{\text{ij}}=\frac{k^i k^j}{k^2}$. The $\Pi_T$ and $\Pi_L$ are obtained by solving Eq.~(\ref{eq:pi}) and Eq.~(\ref{eq:eplt}) simultaneously by making the possible contractions,  
\ba
\Pi_T(K,\nu,T,\mu_q) &=&m_D^2(T,\mu_q)\frac{\omega}{4 k^3} \bigg[2 k (\omega +i \nu )\nn &+& \Big(k^2+(\nu -i \omega )^2\Big)\ln \Big(\frac{\omega+i \nu +k }{\omega+i \nu -k }\Big)\bigg]
\ea
and
\ba
\Pi_L(K,\nu,T,\mu_q) =-m_D^2(T,\mu_q)\frac{\omega ^2}{k^2}\frac{1-\frac{\omega +i \nu}{2 k}\ln \left(\frac{\omega+i \nu +k }{\omega+i \nu -k }\right)}{1-\frac{i \nu}{2 k}  \ln \left(\frac{\omega+i \nu +k }{\omega+i \nu -k }\right)}.\nn
\ea
Similary, the permittivity tensor can be expanded in terms of its longitudinal and transverse components as,
\ba
\epsilon^{\text{ij}}(K,\nu,T,\mu_q)= A^{ij}~\epsilon_T(K,\nu,T,\mu_q)+B^{ij}~\epsilon_L(K,\nu,T,\mu_q),
\label{eq:eplt1}\nn
\ea 
From Eqs.~(\ref{eq:ep}), Eq.~(\ref{eq:eplt}) and Eq.~(\ref{eq:eplt1}), one can obtain the longitudinal and transverse part of the dielectric permittivity respectively as,
\ba
\epsilon_L(K,\nu,T,\mu_q)&=&1- \frac{\Pi_L(K,\nu,T,\mu_q)}{\omega^2}
\ea
and
\ba
\epsilon_T(K,\nu,T,\mu_q)=1- \frac{\Pi_T(K,\nu,T,\mu_q)}{\omega^2},
\ea
these could be further written as,
\ba
\epsilon _L(K,\nu,T,\mu_q)=1+m_D^2(T,\mu_q) \frac{2 k-(\omega +i \nu ) \ln \left(\frac{k+i \nu +\omega }{-k+i \nu +\omega }\right)}{k^2 \left(2 k-i \nu  \ln \left(\frac{k+i \nu +\omega }{-k+i \nu +\omega }\right)\right)}\nn
\ea
and
\ba
\epsilon _T(K,\nu,T,\mu_q)&=&1-\frac{m_D^2(T,\mu_q)}{2~ \omega~k} \bigg[\frac{\omega +i \nu}{k}+\Big(\frac{1}{2}\nn
&-&\frac{(\omega +i \nu )^2}{2k^2}\Big)\ln \left(\frac{k+i \nu +\omega }{-k+i \nu +\omega }\right)\bigg].
\ea
In the next sub-section, we shall discuss the incorporation of hot QCD medium interaction effects using lattice and HTL equations of states (EoSs) along with the inclusion of finite quark chemical potential. 
	 	\begin{figure*}[ht]
		\centering
		\includegraphics[height=5cm,width=8.60cm]{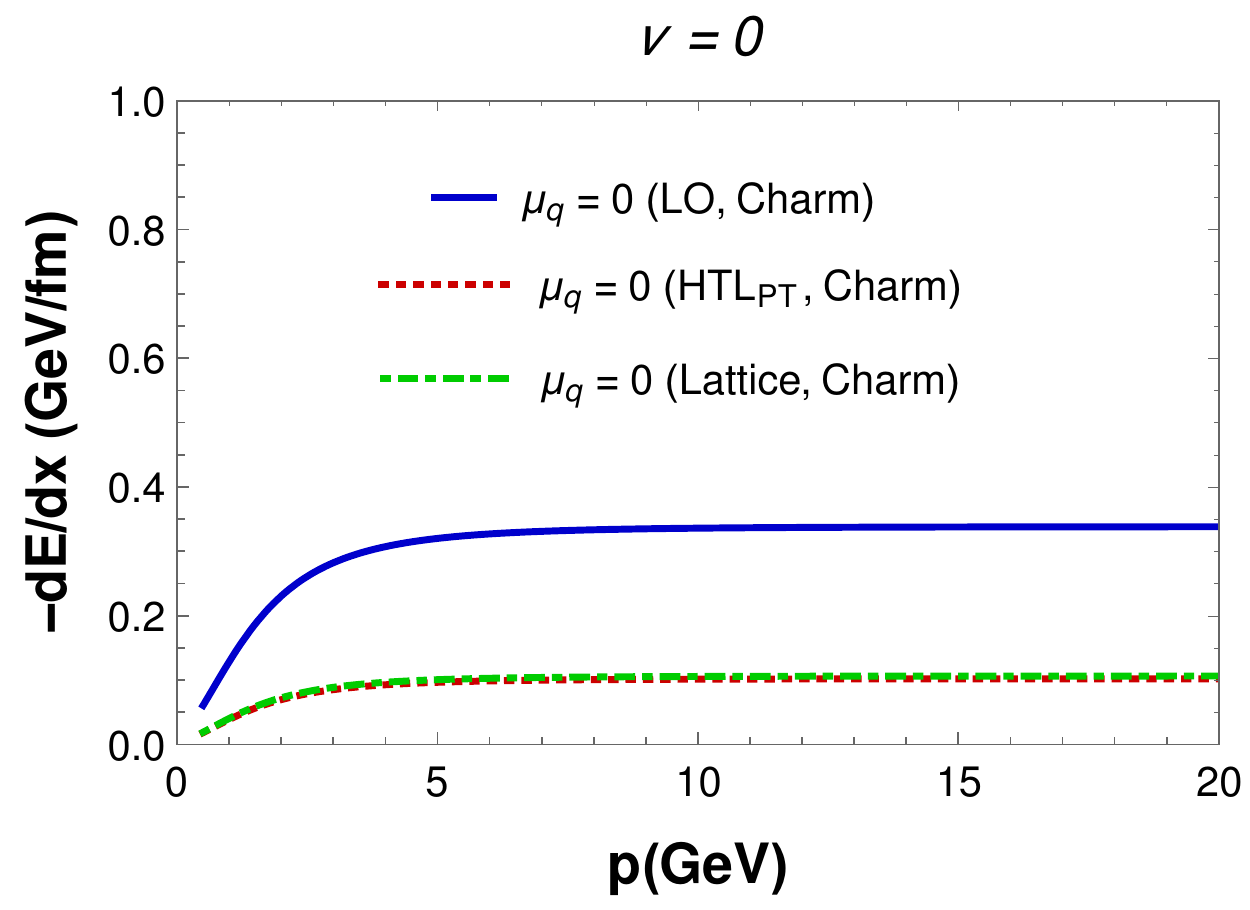}
		\includegraphics[height=5cm,width=8.60cm]{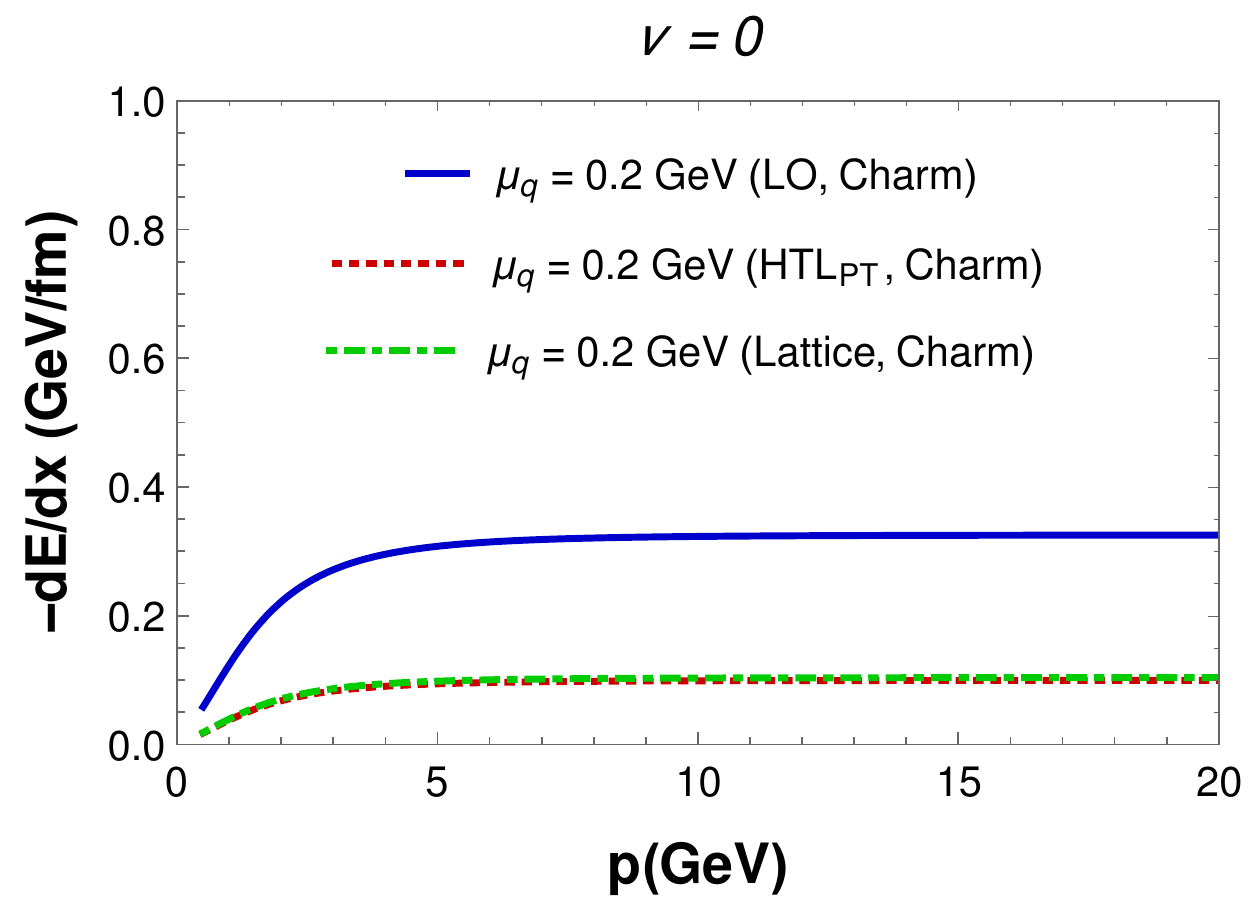}
		\includegraphics[height=5cm,width=8.60cm]{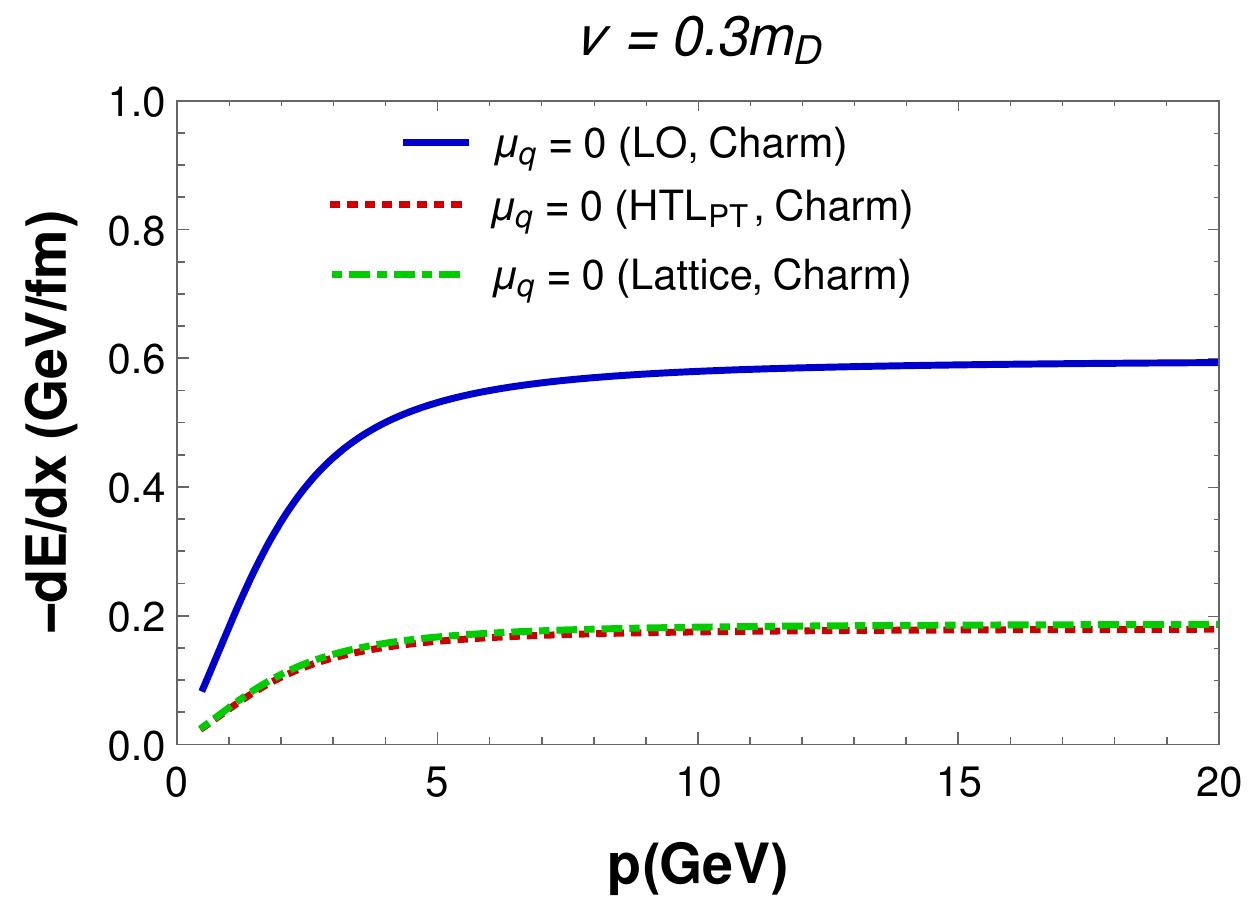}
		\includegraphics[height=5cm,width=8.60cm]{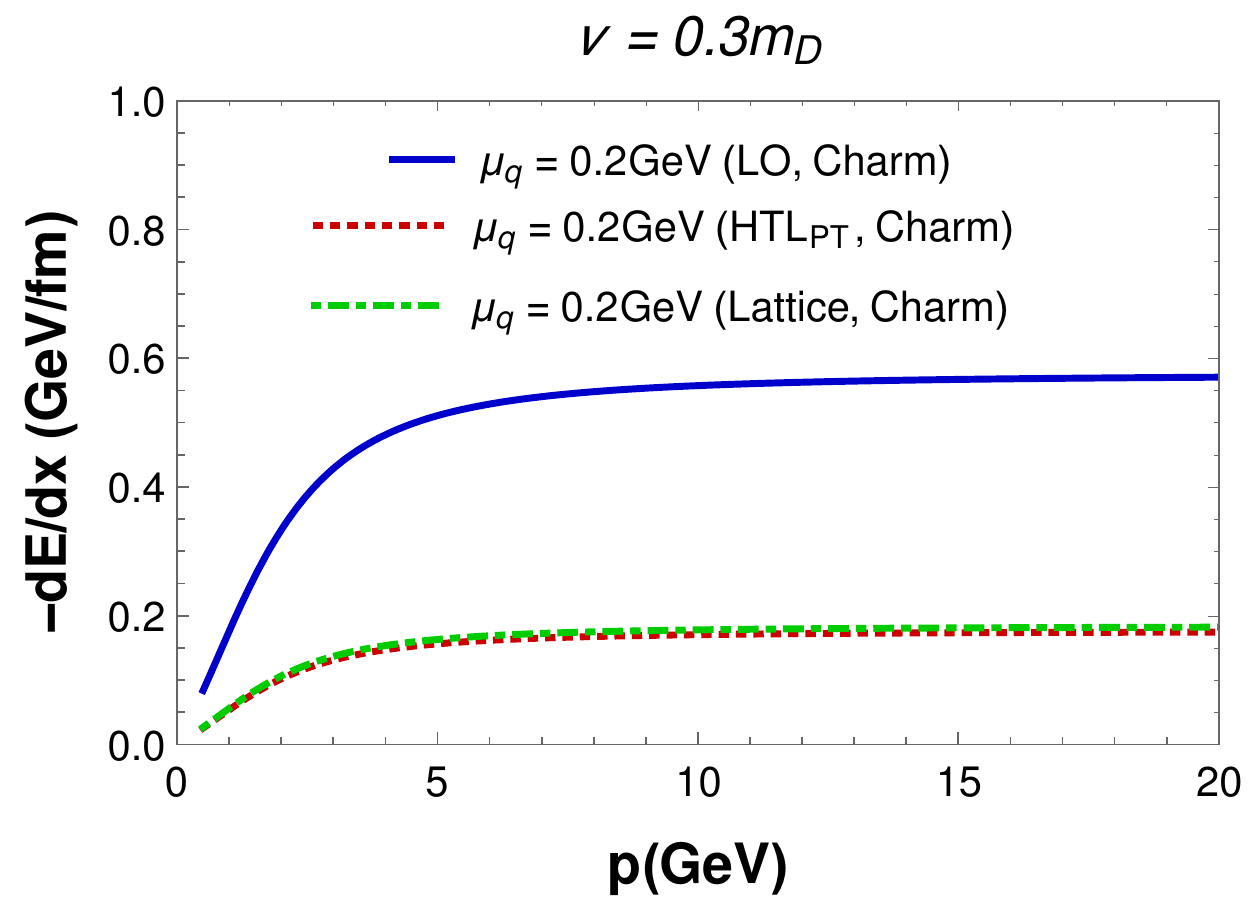}
		\caption{Energy-loss of charm quark ($m=1.8~$GeV) for various EoSs at $T = 2 T_c$ for $\nu=0.0$ (top row), $\nu=0.3 m_D$ (bottom row),  $\mu_q =0.0$ (left column) and $\mu_q =0.2$GeV (right column). }
		\label{fig:charm}
	\end{figure*}
	\subsection{Extended EQPM to incorporate the medium interaction and the finite quark chemical potential }
	To incorporate the small but finite quark chemical potential along with the medium interaction effects, the extended EQPM has been employed (extended with the consideration of finite chemical potential, for details see Ref.~\cite{Mitra:2017sjo}). The EQPM maps the hot QCD medium interaction effects present in the hot QCD EoSs either computed within improved perturbative QCD (pQCD) or lattice QCD simulations into the effective equilibrium distribution functions for the   quasi-partons, $ f_{eq}\equiv \lbrace f_{g}, f_{q},f_{\bar q} \rbrace$, that in turn, describes the strong interaction effects in terms of effective fugacities, $z_{g,q}$~\cite{chandra_quasi1, chandra_quasi2}. The hot QCD EoSs described here in terms of EQPM are the recent ($2+1$)-lattice EoS from hot QCD collaboration ~\cite{bazabov2014} , and the 3-loop HTL perturbative EoS that has recently been computed by N. Haque {\it et,  al.} ~\cite{nhaque, Andersen:2015eoa} which agrees reasonably well with the recent lattice results~\cite{bazabov2014, fodor2014}. The EQPM recently applied to study the various aspects of hot QCD medium ~\cite{ Jamal:2017dqs, Kurian:2019nna, Jamal:2018mog, Agotiya:2016bqr, Kumar:2017bja, YousufJamal:2018ucf, Jamal:2017ygv}.
	Next, in the absence of chemical potential the quark and gluon distribution functions are defined as,  
	\ba
	\label{eq1}
	f_{g/q}= \frac{z_{g/q}\exp[-\beta E_p]}{\bigg(1\mp z_{g/q}\exp[-\beta E_p]\bigg)},
	\ea
	where, $E_{p}=|{\bf p}|$ for the gluons and, $\sqrt{|{\bf p}|^2+m_q^2}$ for the quark ($m_q$, denotes the mass of the quarks).	In this case, we have $f_{q}=f_{{\bar q}}$. The fugacity parameter, $z_{g/q}\rightarrow 1$ as temperature $T\rightarrow \infty$.
	Since, the model is valid only in the deconfined phase of QCD (beyond $T_c$, $T_c$ being the critical temperature), the mass of the light quarks can be neglected. In the present case, at finite quark chemical potential the momentum distribution of quark and anti-quark get modify as,
	\ba
	\label{eq2}
	f_{q}&=& \frac{z_{q}\exp[-\beta ( E_p-\mu_q)]}{\bigg(1+ z_{q}\exp[-\beta ( E_p-\mu_q)]\bigg)},\nn f_{\bar q}&=& \frac{z_{q}\exp[-\beta ( E_p+\mu_q)]}{\bigg(1+ z_{q}\exp[-\beta ( E_p+\mu_q)]\bigg)}.
	\ea
		Solving Eq.~(\ref{eq:md}), considering the distribution functions given in Eq.~(\ref{eq2}) within the limit, $\frac{\mu^2_q}{T^2}\ll1$ we obtained,
	 \ba
	m^{2, EoS}_D(T,\mu_q) &=& 4\pi \alpha_s(T,\mu_q)T^2 \bigg[\Big(\frac{2N_c}{\pi^2}PolyLog[2,z_g^{EoS}]\nn
	&-&\frac{2N_f}{\pi^2}PolyLog[2,-z_q^{EoS}]\Big)+\frac{\mu_q^2}{T^2}\frac{N_f}{\pi^2}\frac{z_q^{EoS}}{1+z_q^{EoS}}
	\bigg],\nn
	\label{eq:mde}
		\ea
  In the high temperature limit, $z_{g,q}\rightarrow 1$ and Eq.~(\ref{eq:mde}) reduces to leading order HTL expression,
		\ba
	m^{2}_D(T,\mu_q) &=& 4\pi \alpha_s(T,\mu_q)T^2 \bigg[\frac{N_c}{3}+\frac{N_f}{6}
	+\frac{\mu_q^2}{T^2}\frac{N_f}{2\pi^2}
	\bigg].
	\label{eq:mdh}
	\ea
	 From Eq.~(\ref{eq:mde}) and Eq.~(\ref{eq:mdh}), one can obtain the effective coupling due to medium interaction effects through various EoSs at finite chemical potential as,
	  \ba
	\alpha_s^{EoS}(T,\mu_q) &=&\frac{\alpha_s(T,\mu_q)}{\Big(\frac{N_c}{3}+\frac{N_f}{6}\Big)
		+\frac{\mu_q^2}{T^2}\frac{N_f}{2\pi^2}}  \bigg[\Big(\frac{2N_c}{\pi^2}\text {PolyLog}[2,z_g^{EoS}]\nn
	 &-&\frac{2N_f}{\pi^2}\text {PolyLog}[2,-z_q^{EoS}]\Big)
	 +\frac{\mu_q^2}{T^2}\frac{N_f}{\pi^2}\frac{z_q^{EoS}}{1+z_q^{EoS}}
	 \bigg].\nn
	 \label{eq:ec}
	 \ea
	\begin{figure*}[ht]
 	\centering
 	\includegraphics[height=5cm,width=8.60cm]{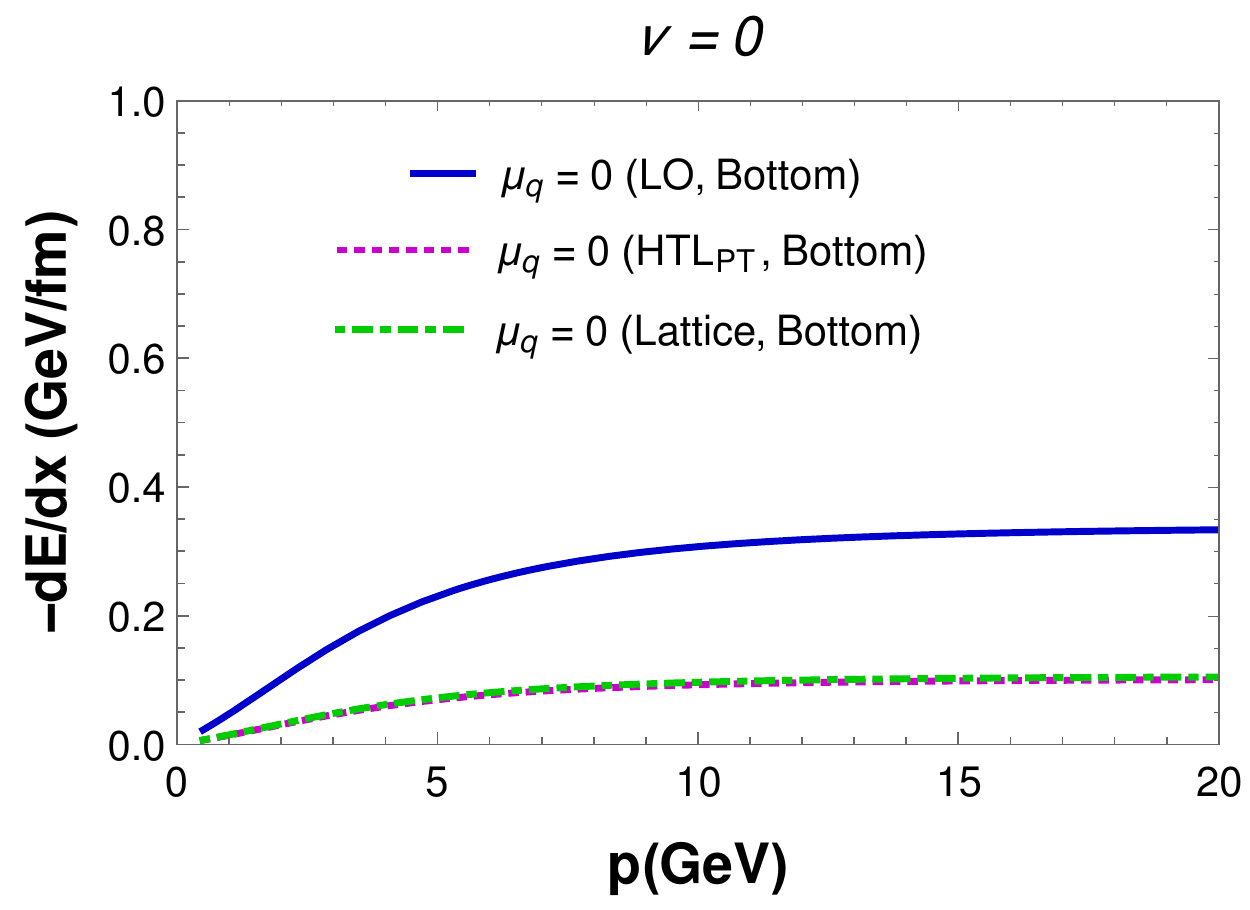}
 	\includegraphics[height=5cm,width=8.60cm]{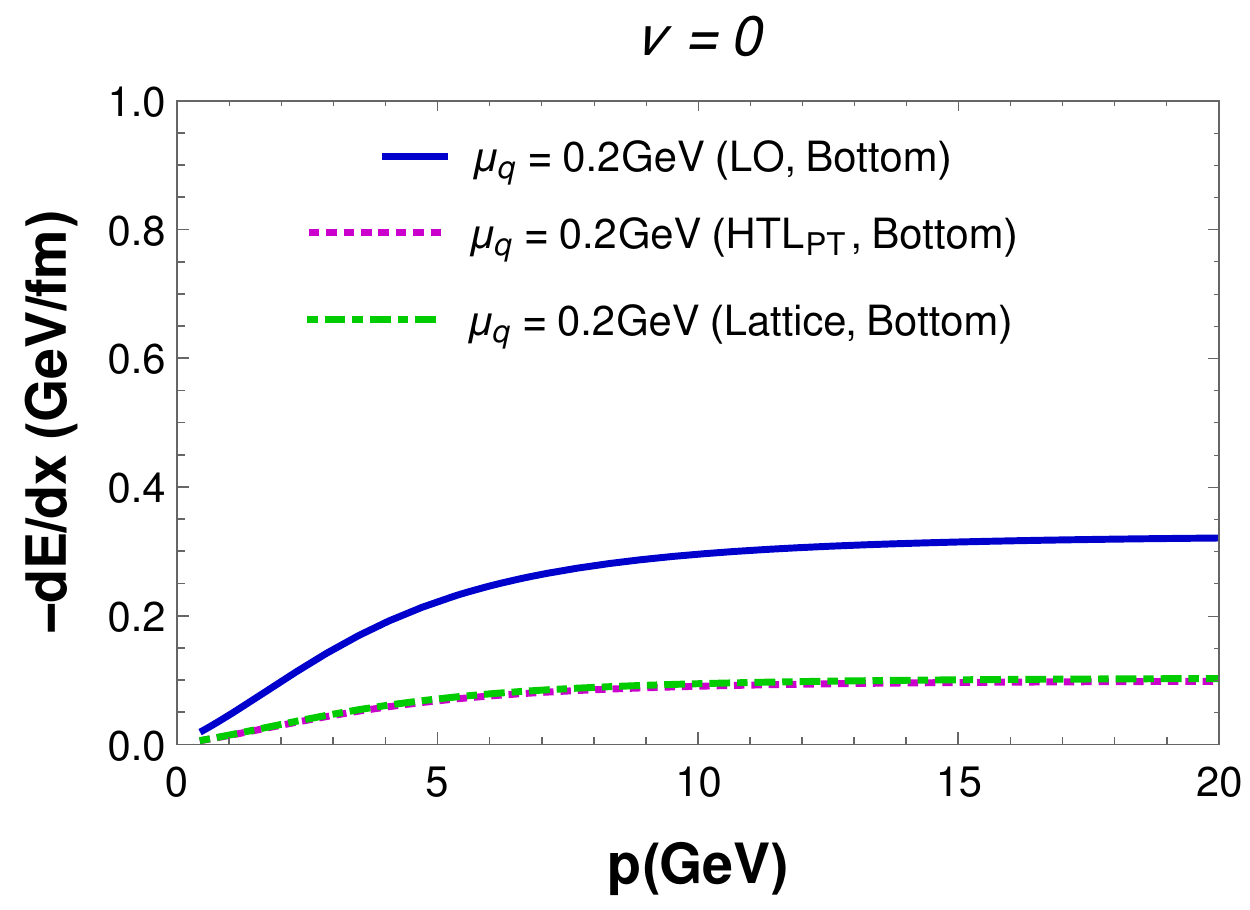}
 	\includegraphics[height=5cm,width=8.60cm]{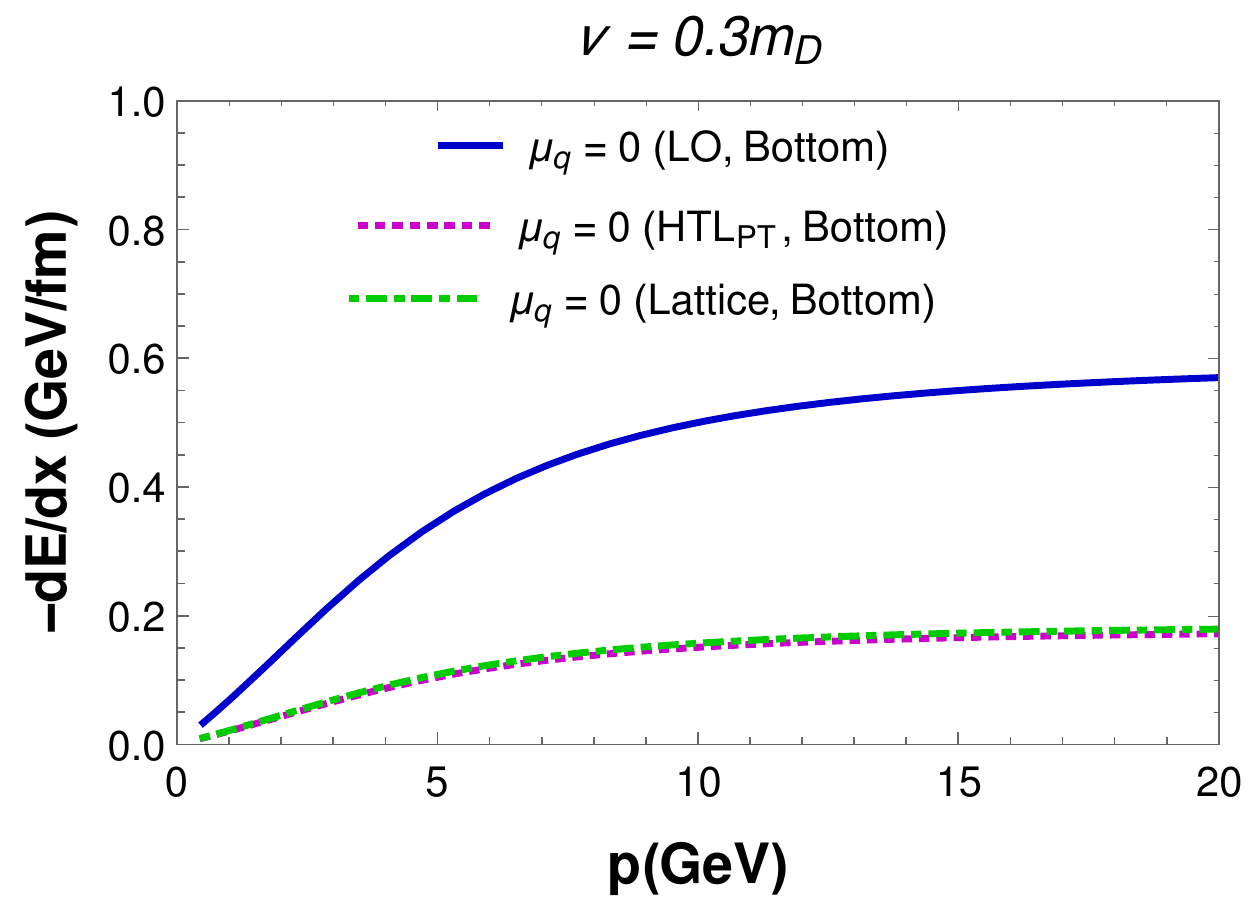}
 	\includegraphics[height=5cm,width=8.60cm]{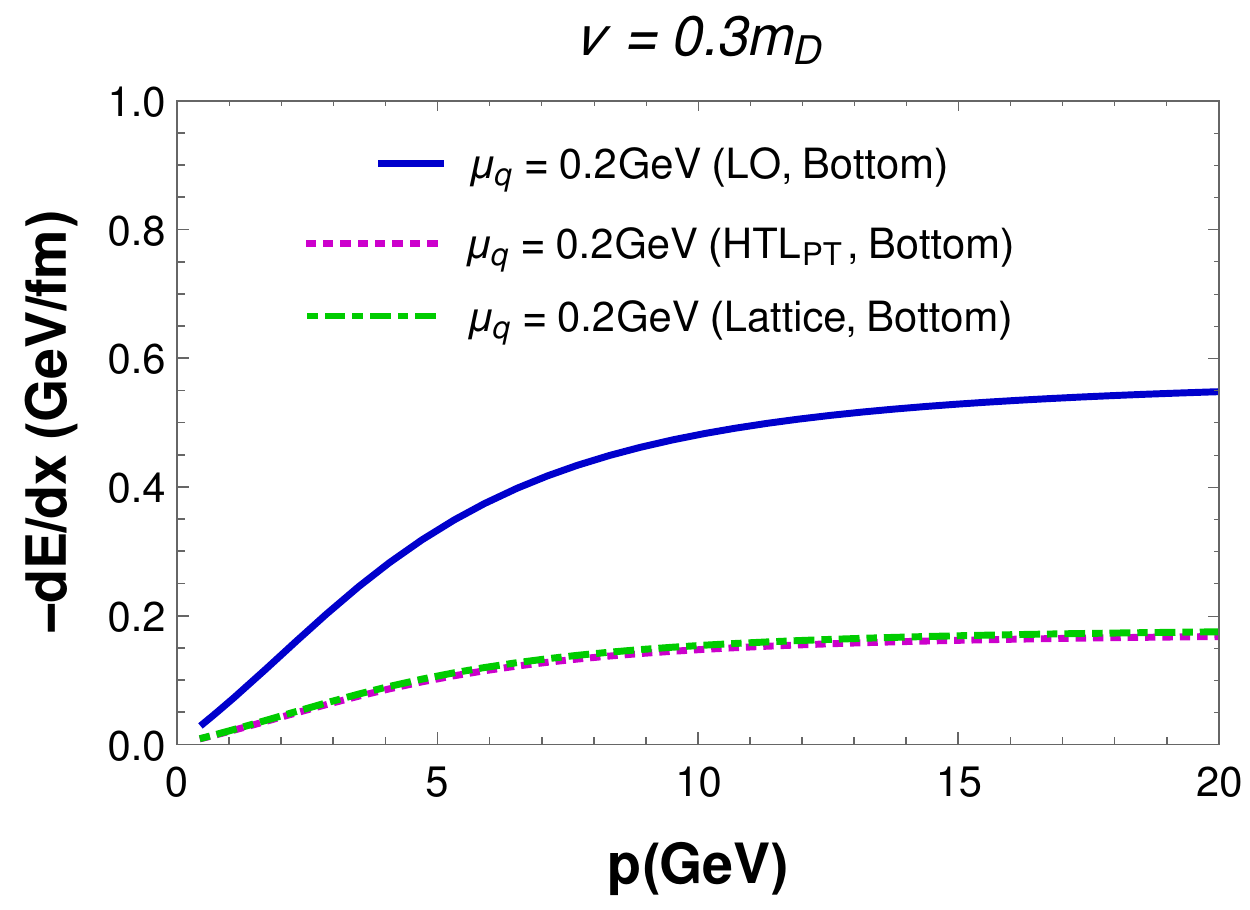}
 	\caption{Energy-loss of bottom quark ($m=4.5~$GeV) for various EoSs at $T = 2 T_c$ for $\nu=0.0$ (top row), $\nu=0.3 m_D$ (bottom row),  $\mu_q =0.0$ (left column) and $\mu_q =0.2$GeV (right column). }
 	\label{fig:bottom}
 \end{figure*}	 

Next, we shall discuss the various results of energy-loss of charm and bottom quarks plotted against their momenta for various EoSs at different collision frequency and chemical potential.

	\section{Results and discussion}
	\label{el:RaD}
The formalism for the energy-loss of heavy quarks moving in the presence of interacting isotropic collisional hot QCD/QGP medium has been extended by considering the effects of small but finite quark chemical potential. To obtain the corresponding results, Eq.(\ref{eq:de}) have been derived and solved numerically. To perform the numerical integration, the lower and upper limits, respectively  have been taken as $k^{EoSs}_{0}=0$ and $k^{EoSs}_{\infty}\sim m^{EoSs}_{D}(T)$ for each EoS. Here, we worked at temperature, $T=2~T_c$ where, $T_c=0.17 $~GeV. The results obtained for the ideal case (or the leading order (LO)) are compared with the non-ideal cases ( $(2+1)-$ lattice EoS and 3-loop HTL EoS denoted as LB and  $\mathrm{HTL_{PT}}$, respectively in the plots). The energy-loss of the charm and bottom quarks have been plotted against their momenta at different collision frequency, $\nu$ and chemical potential, $\mu_q$ as shown in the various figures. 

	\begin{figure*}[ht]
	\centering
	\includegraphics[height=5cm,width=8.60cm]{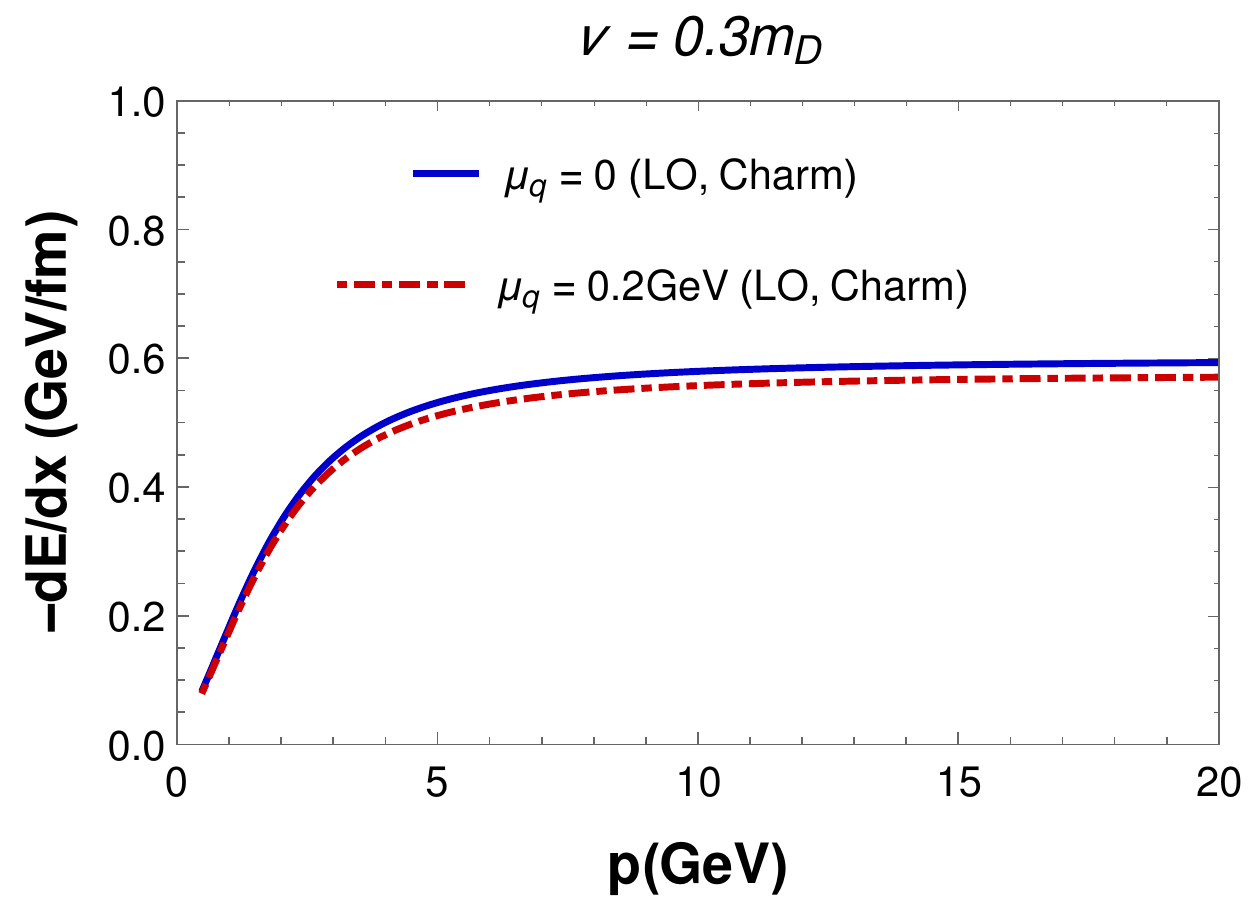}
	\includegraphics[height=5cm,width=8.60cm]{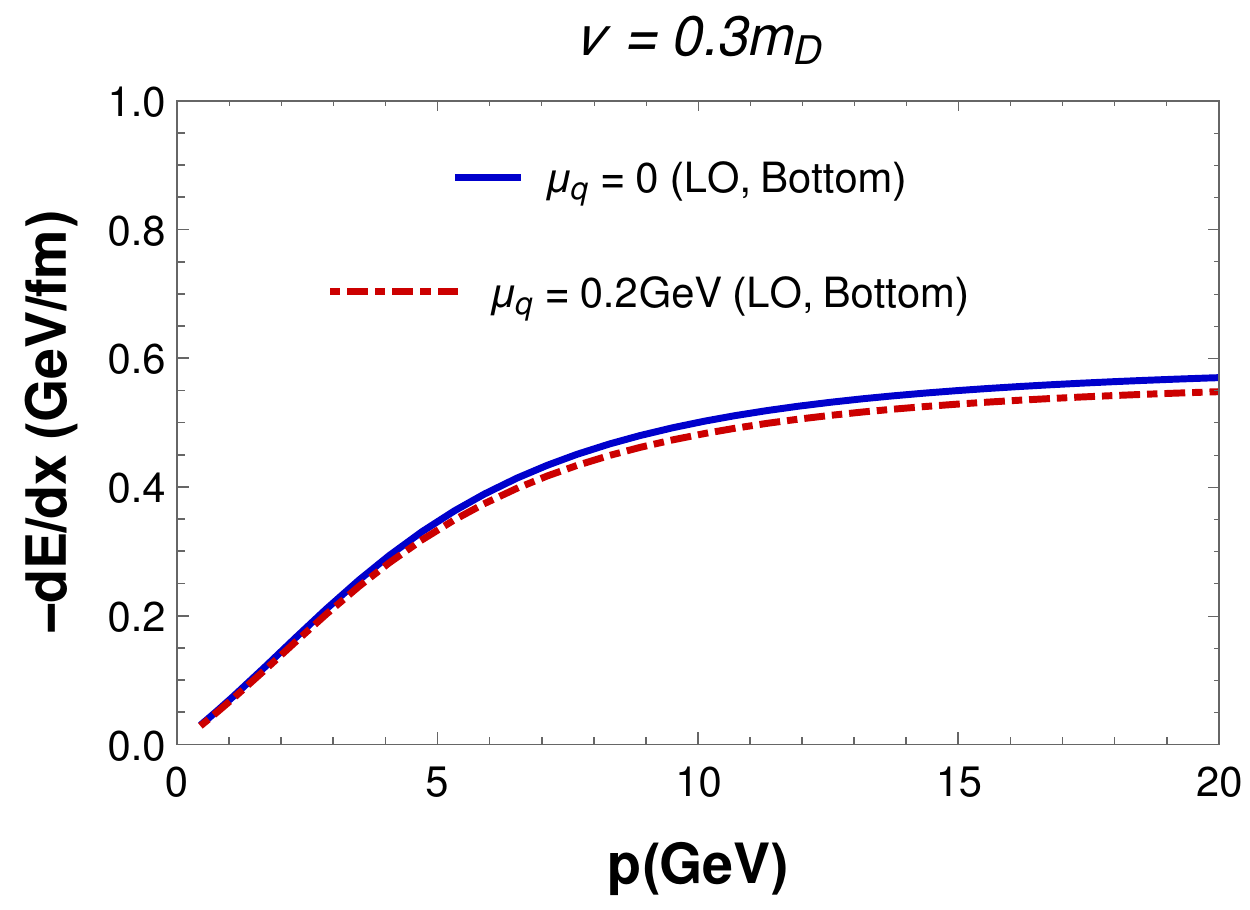}
	\caption{Energy-loss of charm quark (left column) and bottom quark  (right column)  at $\nu = 0.3 m_{D}$, $T = 2 T_c$ and different values of $\mu_q$.}
	\label{fig:CBMU}
\end{figure*}
\begin{figure*}[ht] 
	\centering
	\includegraphics[height=5cm,width=8.60cm]{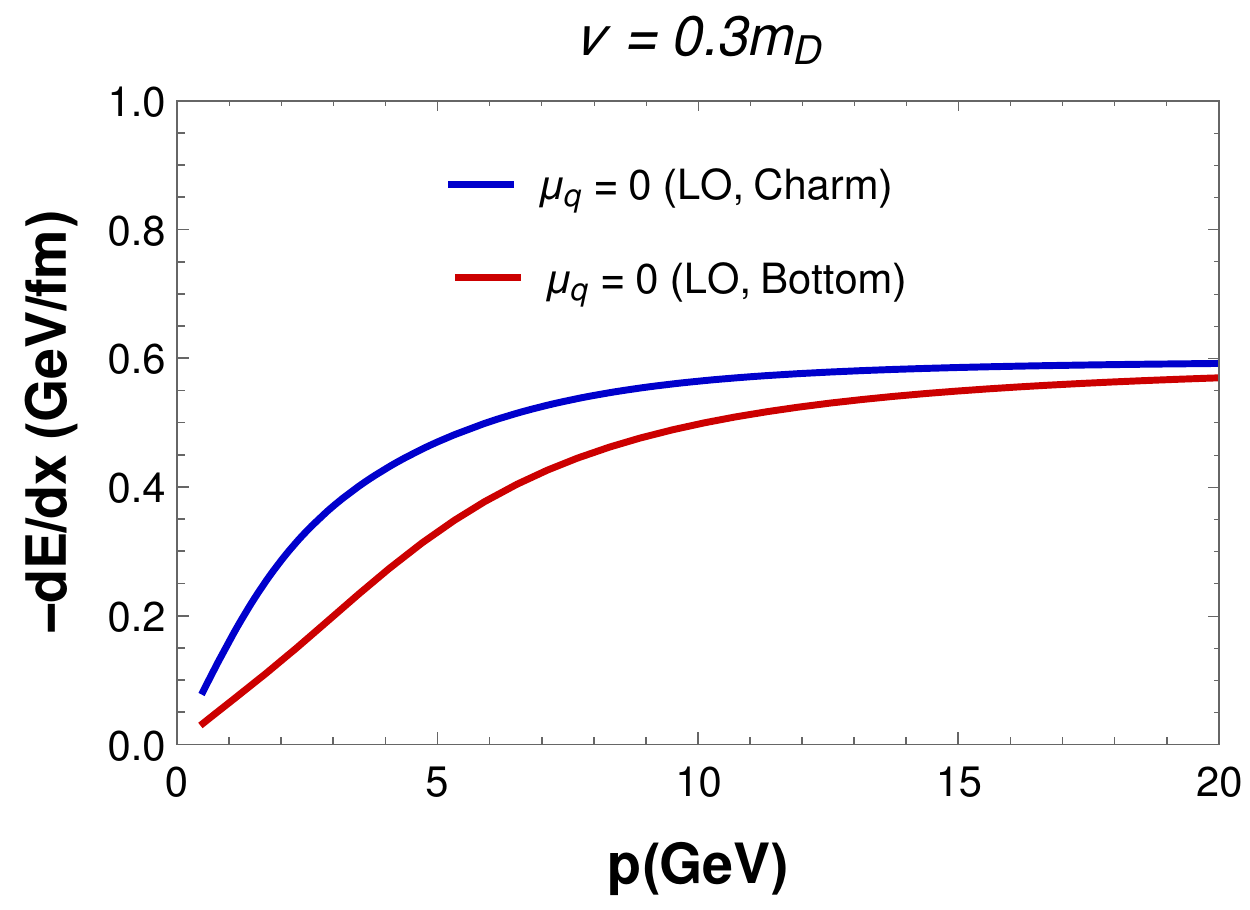}
	\includegraphics[height=5cm,width=8.60cm]{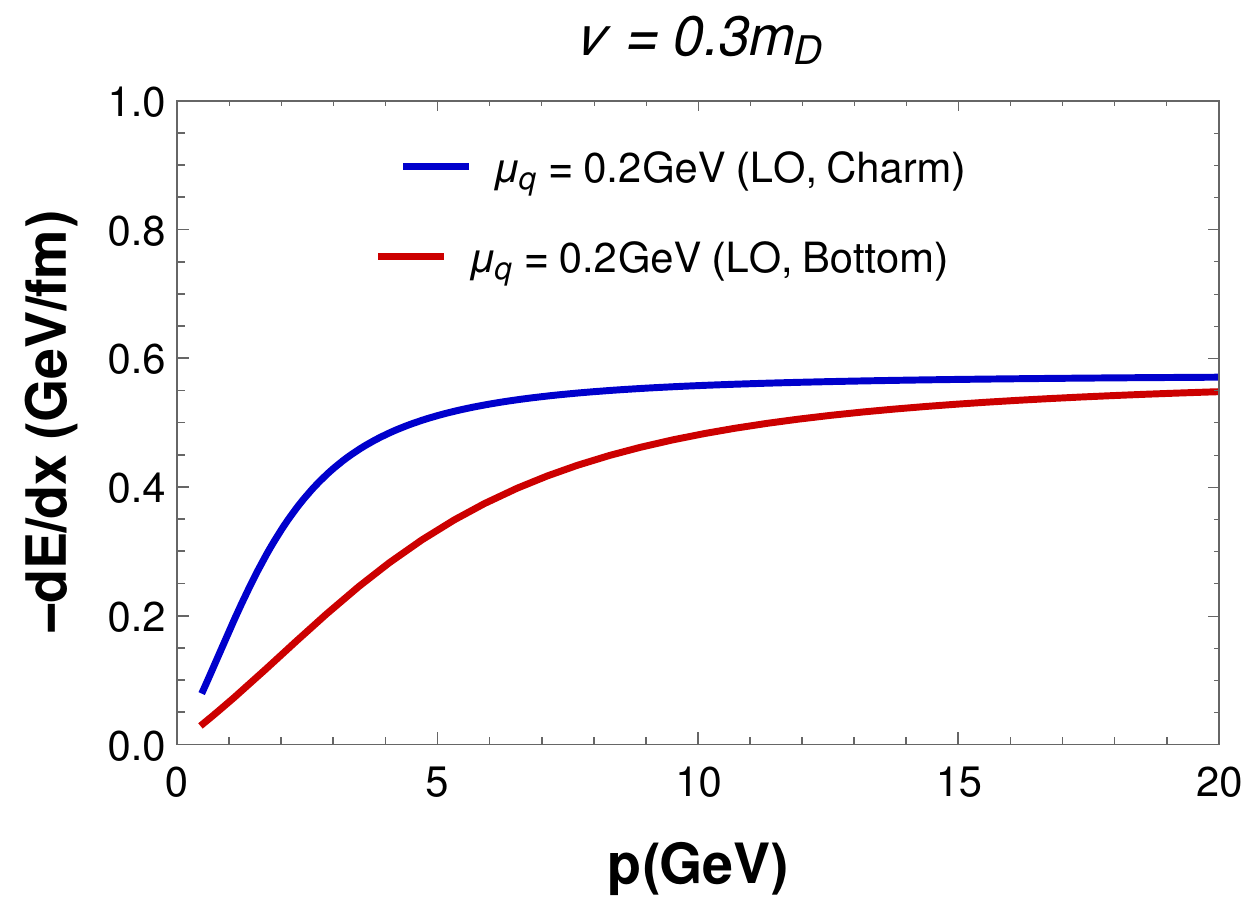}
	\caption{Comparison between the energy-loss of charm and bottom quarks at $\mu_q=0$  (left column), $\mu_q=0.2$GeV  (right column) and $T = 2 T_c$. }
	\label{fig:CB}
\end{figure*}

In Fig.~\ref{fig:charm}, a comparison has been shown for the energy-loss of charm quark at different values of $\nu$ and $\mu_q$  in a matrix form. The top row is having results for the collisionless case whereas, for the collision frequency, $\nu=0.3 ~m_D$ the results are shown in the bottom row. The effect of finite chemical potential is shown in columns. The left column is plotted for vanishing chemical potential while the right column is showing the results for small but finite chemical potential, $\mu_q= 0.2~$GeV. One can observe that with the increase in chemical potential, energy-loss decreases whereas with the increase in collision frequency it increases. The results for bottom quark have been shown in Fig.~\ref{fig:bottom}. It has been noticed that the energy-loss of bottom quark follows a similar pattern as charm quark with slightly different numbers. The medium interaction effects that are entering through the non-ideal EoSs are found to suppress the energy-loss in both the cases ( charm and bottom ). 

In Fig.~\ref{fig:CBMU}, only the LO results has been shown at fixed collision frequency, $\nu =0.3~m_D$ to make the effects of finite chemical potential more visible. As we have considered small chemical potential, the results are found to be closer to the vanishing case. Although, one can observe that the heavy quarks lose less energy at $\mu_q=0.2~$GeV as compared to  $\mu_q=0$.
In Fig.~\ref{fig:CB}, a comparison has been shown between the energy-loss of charm and bottom quarks at fixed collision frequency, $\nu =0.3~m_D$ and at $\mu_q=0$ and $\mu_q=0.2~$GeV. Again, only the LO case has been considered here. It can be observed that the bottom quark loses less energy as compared to the charm quark in both the cases at $\mu_q=0$ (left panel) and $\mu_q=0.2~$GeV (right panel). At vanishing chemical potential as well as at finite chemical potential, the bottom quark is observed to lose less energy.
This result preserves the fact that heavy particle loses less energy while moving in a medium than the lighter one, given the same conditions and hence, thermalizes late comparatively.

Here, the main aim was to incorporate the finite chemical potential in the analysis, therefore, to avoid the bulk we did not show more plots on the variation of collision frequency and EoSs. For more details regarding the effects of collision frequency and EoSs, one can go through the Ref.~\cite{Jamal:2019svc}.  Furthermore, the results observed here are found to be consistent with the partial results available in the literature ~\cite{Han:2017nfz, Berrehrah:2015ywa}.

\section{Summary and future aspects}
\label{el:SaF}
We have discussed the energy-loss of the heavy quarks traversing through the interacting isotropic collisional hot QCD medium produced in the relativistic heavy-ion collision experiments considering the finite chemical potential.  To do so, the analytic expression for the energy-loss of heavy quarks within the effective kinetic theory approach has been obtained in term of dielectric permittivity of the medium.
To incorporate the collision effects, the BGK-kernel has been considered. Whereas the non- ideal interaction effects, as well as the small but finite chemical potential, have been taken into account through the equilibrium particle distribution functions by employing extended EQPM.
It has been observed that the energy-loss initially increases with the quark momentum and then saturates (as moves above the corresponding quark mass). The energy-loss is found to increase with the increase in collision frequency whereas, an opposite behaviour is observed with the finite chemical potential.
Apart from that, the charm quark is found to lose more energy than the bottom quark given the same collision frequency and chemical potential. 

An immediate future extension of the current work would be the study of magnetic field effects, momentum anisotropy as well as viscous effects considering the hot QCD/QGP medium as interacting. Since, the results presented here are completely theoretical, in the very near future, we would like to investigate the nuclear modification factor to map them with the experimental observations.
    
\section{Acknowledgements}
M. Y. Jamal acknowledges NISER Bhubaneswar for providing postdoctoral position. We would like to acknowledge people of INDIA for their generous support for the research in fundamental sciences in the country.


\begin{thebibliography}{0}
	\bibitem{Mitra:2017sjo} 
S.~Mitra and V.~Chandra,
Phys.\ Rev.\ D {\bf 96}, 094003 (2017).

\bibitem{chandra_quasi1}
V. Chandra, R. Kumar, V. Ravishankar, Phys. Rev.  C {\bf 76}, 054909 (2007);
[Erratum: Phys. Rev. C {\bf 76}, 069904 (2007)];
V. Chandra, A. Ranjan, V. Ravishankar, Eur. Phys. J. A {\bf 40}, 109-117 (2009).


\bibitem{chandra_quasi2}
V. Chandra, V. Ravishankar, Phys. Rev.  D {\bf 84}, 074013 (2011). 


\bibitem{bjorken1982energy}
Bjorken, James D, Fermilab preprint 82/59-THY, 1982, unpublished. 

\bibitem{Thoma:1990fm}
M.~H.~Thoma and M.~Gyulassy,
Nucl.\ Phys.\ B {\bf 351}, 491 (1991).

\bibitem{Braaten:1991jj}
E.~Braaten and M.~H.~Thoma,
Phys.\ Rev.\ D {\bf 44}, 1298 (1991); Phys.\ Rev.\ D {\bf 44},  R2625  (1991).


\bibitem{Mrowczynski:1991da}
S.~Mrowczynski,
Phys.\ Lett.\ B {\bf 269},383 (1991).


\bibitem{Thomas:1991ea}
M.~H.~Thoma,
Phys.\ Lett.\ B {\bf 273}, 128 (1991).

\bibitem{Koike:1992xs}
Y.~Koike and T.~Matsui,
Phys.\ Rev.\ D {\bf 45}, 3237 (1992).

\bibitem{Romatschke:2004au}
P.~Romatschke and M.~Strickland,
Phys.\ Rev.\ D {\bf 69},065005 (2004) ; Phys.\ Rev.\ D {\bf 71}, 125008 (2005).

\bibitem{Baier:2008js}
R.~Baier and Y.~Mehtar-Tani,
Phys.\ Rev.\ C {\bf 78}, 064906 (2008).


\bibitem{Carrington:2015xca}
M.~E.~Carrington, K.~Deja and S.~Mrowczynski,
Phys.\ Rev.\ C {\bf 92}, 044914 (2015); Phys.\ Rev.\ C {\bf 95},  024906 (2017).

\bibitem{Baier:2000mf}
R.~Baier, D.~Schiff and B.~G.~Zakharov,
Ann.\ Rev.\ Nucl.\ Part.\ Sci.\  {\bf 50}, 37 (2000).

\bibitem{Jacobs:2004qv}
P.~Jacobs and X.~N.~Wang,
Prog.\ Part.\ Nucl.\ Phys.\  {\bf 54}, 443 (2005).

\bibitem{Armesto:2011ht}
N.~Armesto {\it et al.},
Phys.\ Rev.\ C {\bf 86},064904 (2012). 

\bibitem{Majumder:2010qh}
A.~Majumder and M.~Van Leeuwen,
Prog.\ Part.\ Nucl.\ Phys.\  {\bf 66}, 41 (2011).

\bibitem{Mustafa:1997pm}
M.~G.~Mustafa, D.~Pal, D.~K.~Srivastava and M.~Thoma,
Phys.\ Lett.\ B {\bf 428}, 234 (1998).

\bibitem{Dokshitzer:2001zm} 
Y.~L.~Dokshitzer and D.~E.~Kharzeev,
Phys.\ Lett.\ B {\bf 519}, 199 (2001)

\bibitem{Djordjevic:2003zk}
M.~Djordjevic and M.~Gyulassy,
Nucl.\ Phys.\ A {\bf 733}, 265 (2004).

\bibitem{Wicks:2007am}
S.~Wicks, {\it et.al.,}  
Nucl.\ Phys.\ A {\bf 783} (2007) 493; Nucl.\ Phys.\ A {\bf 784},  426 (2007).

\bibitem{Abir:2011jb}
R.~Abir,{\it et.al.,}
Phys.\ Rev.\ D {\bf 85} (2012) 054012; Phys.\ Lett.\ B {\bf 715}, 183 (2012).

\bibitem{Jeon:2003gi}
S.~Jeon and G.~D.~Moore,
Phys.\ Rev.\ C {\bf 71}, 034901 (2005).

	\bibitem{Gyulassy:1999zd}  M.~Gyulassy, P.~Levai and I.~Vitev,
Nucl.\ Phys.\ B {\bf 571}, 197 (2000).

\bibitem{Zakharov:2000iz}  B.~G.~Zakharov,
JETP Lett.\  {\bf 73},49 (2001).

\bibitem{Djordjevic:2008iz}
M.~Djordjevic and U.~W.~Heinz,
Phys.\ Rev.\ Lett.\  {\bf 101}, 022302 (2008).

\bibitem{Baier:2001yt}
R.~Baier, Y.~L.~Dokshitzer, A.~H.~Mueller and D.~Schiff,
JHEP {\bf 0109}, 033 (2001).

\bibitem{Qin:2007rn}
G.~Y.~Qin, J.~Ruppert, C.~Gale, S.~Jeon, G.~D.~Moore and M.~G.~Mustafa,
Phys.\ Rev.\ Lett.\  {\bf 100}, 072301 (2008).

\bibitem{Cao:2013ita}
S.~Cao, G.~Y.~Qin and S.~A.~Bass,
Phys.\ Rev.\ C {\bf 88}, 044907 (2013).

\bibitem{Mustafa:2003vh}
M.~G.~Mustafa and M.~H.~Thoma,
Acta Phys.\ Hung.\ A {\bf 22} (2005) 93;   Phys.\ Rev.\ C {\bf 72}, 014905 (2005).

\bibitem{DuttMazumder:2004xk}
A.~K.~Dutt-Mazumder, J.~e.~Alam, P.~Roy and B.~Sinha,
Phys.\ Rev.\ D {\bf 71},094016 (2005). 

\bibitem{Meistrenko:2012ju}
A.~Meistrenko, A.~Peshier, J.~Uphoff and C.~Greiner,
Nucl.\ Phys.\ A {\bf 901}, 51 (2013).

\bibitem{Burke:2013yra}
K.~M.~Burke {\it et al.} [JET Collaboration],
Phys.\ Rev.\ C {\bf 90},  014909 (2014).


\bibitem{Peigne:2007sd}
S.~Peigne and A.~Peshier,
Phys.\ Rev.\ D {\bf 77} (2008) 014015;  Phys.\ Rev.\ D {\bf 77}, 114017 (2008).

\bibitem{Neufeld:2014yaa}
R.~B.~Neufeld, I.~Vitev and H.~Xing,
Phys.\ Rev.\ D {\bf 89} (2014),  096003 (2014).

\bibitem{Chakraborty:2006db}
P.~Chakraborty, M.~G.~Mustafa and M.~H.~Thoma,
Phys.\ Rev.\ C {\bf 75}, 064908 (2007).

\bibitem{Adil:2006ei}
A.~Adil, M.~Gyulassy, W.~A.~Horowitz and S.~Wicks,
Phys.\ Rev.\ C {\bf 75}, 044906 (2007).

\bibitem{Peigne:2005rk}
S.~Peigne, P.~B.~Gossiaux and T.~Gousset,
JHEP {\bf 0604}, 011(2006).

\bibitem{Dusling:2009jn} 
K.~Dusling and I.~Zahed,
Nucl.\ Phys.\ A {\bf 833}, 172 (2010).


\bibitem{Cho:2009ze} 
S.~Cho and I.~Zahed,
Phys.\ Rev.\ C {\bf 82}, 064904 (2010).

\bibitem{Jiang:2014oxa}
B.~F.~Jiang, D.~Hou and J.~R.~Li,
J.\ Phys.\ G {\bf 42},  085107 (2015).

\bibitem{Jiang:2016duz}
B.~f.~Jiang, D.~f.~Hou and J.~r.~Li,
Nucl.\ Phys.\ A {\bf 953}, 176 (2016).

\bibitem{Elias:2014hua}
M.~Elias, J.~Peralta-Ramos and E.~Calzetta,
Phys.\ Rev.\ D {\bf 90}, 014038 (2014).


\bibitem{Han:2017nfz}
C.~Han, D.~f.~Hou, B.~f.~Jiang and J.~r.~Li,
Eur.\ Phys.\ J.\ A {\bf 53}, 205 (2017).

\bibitem{Fadafan:2008gb} 
K.~B.~Fadafan,
JHEP {\bf 0812}, 051 (2008).

\bibitem{Fadafan:2008uv} 
K.~Bitaghsir Fadafan,
Eur.\ Phys.\ J.\ C {\bf 68}, 505 (2010).

\bibitem{Fadafan:2012qu} 
K.~B.~Fadafan and H.~Soltanpanahi,
JHEP {\bf 1210}, 085 (2012).

\bibitem{Jamal:2019svc} 
M.~Yousuf Jamal and V.~Chandra,
Eur.\ Phys.\ J.\ C {\bf 79}, 761 (2019).

\bibitem{Srivastava:2010xa} 
P.~K.~Srivastava, S.~K.~Tiwari and C.~P.~Singh,
Phys.\ Rev.\ D {\bf 82}, 014023 (2010).

\bibitem{Jiang:2016dkf}
Bing-feng Jiang, De-fu Hou and Jia-rong Li, Phys. Rev. D {\bf 94}, 074026 (2016).

\bibitem{Schenke:2006xu}
Bjoern Schenke, Michael Strickland, Carsten Greiner, Markus H. Thoma, Phys. Rev. D {\bf 73}, 125004 (2006).	

\bibitem{Mrowczynski:1993qm}
S.~Mrowczynski, Phys.\ Lett.\ B {\bf 314}, 118 (1993).		

\bibitem{Romatschke:2003ms}
P. Romatschke and M.Strickland, Phys. Rev. D {\bf 68}, 036004 (2003).

\bibitem{Bhatnagar:1954} 
P. L. Bhatnagar, E. P. Gross, and M. Krook, Phys. Rev. {\bf 94}, 511 (1954).

\bibitem{bazabov2014} A. Bazabov {\it et, al.},  Phys.  Rev  D {\bf 90}, 094503 (2014).

\bibitem{nhaque}
N. Haque, A.  Bandyopadhyay, J. O. Andersen, Munshi G. Mustafa, M. Strickland and Nan Su, 
JHEP {\bf 1405}, 027 (2014).

\bibitem{Andersen:2015eoa}
J. O. Andersen, N. Haque, M. G. Mustafa and M. Strickland,
Phys. Rev. D {\bf 93},  054045  (2016)

\bibitem{fodor2014}
Szabocls Borsanyi  {\it et. al}, Phys. Lett.  B {\bf 370}, 99-104 (2014).

\bibitem{Jamal:2017dqs} M.~Yousuf, S.~Mitra and V.~Chandra,
Phys.\ Rev.\ D {\bf 95},  094022 (2017).

\bibitem{Kurian:2019nna} 
M.~Kurian, S.~K.~Das and V.~Chandra,
arXiv:1907.09556 [nucl-th].

\bibitem{Jamal:2018mog} 
M.~Y.~Jamal, I.~Nilima, V.~Chandra and V.~K.~Agotiya,
Phys.\ Rev.\ D {\bf 97}, no. 9, 094033 (2018).

\bibitem{Agotiya:2016bqr} 
V.~K.~Agotiya, V.~Chandra, M.~Y.~Jamal and I.~Nilima,
Phys.\ Rev.\ D {\bf 94}, no. 9, 094006 (2016).

\bibitem{Kumar:2017bja}
A.~Kumar, M.~Y.~Jamal, V.~Chandra and J.~R.~Bhatt,
Phys.\ Rev.\ D {\bf 97} (2018) no.3,  034007


\bibitem{YousufJamal:2018ucf} 
M.~Yousuf Jamal, S.~Mitra and V.~Chandra,
Springer Proc.\ Phys.\  {\bf 203}, 117 (2018).

\bibitem{Jamal:2017ygv} 
M.~Yousuf, Jamal, S.~Mitra and V.~Chandra,
arXiv:1706.02995 [nucl-th].

\bibitem{Berrehrah:2015ywa} 
H.~Berrehrah, E.~Bratkovskaya, W.~Cassing, P.~B.~Gossiaux and J.~Aichelin,
Phys.\ Rev.\ C {\bf 91}, 054902 (2015).
	\end{thebibliography}
\end{document}